\documentclass[conference]{IEEEtran}
\newcommand{\mbf}[1]{\bm{#1}}

\newcommand{\rth}{^\text{th}}

\newcommand{\eye}{\mbf{I}}
\newcommand{\complex}{\mathbb{C}}

\newcommand{\numBsAntenna}{M}
\newcommand{\numUeAntenna}{K}

\newcommand{\subcarrierIdx}{n}
\newcommand{\channelMtx}[1]{\mbf{H}_{#1}}
\newcommand{\channelEstimate}[1]{\mbf{\widehat{H}}_{#1}}
\newcommand{\receivedSignal}[1]{\mbf{Y}_{#1}}
\newcommand{\ulTxPower}{\eta}
\newcommand{\ulPilotSignal}[1]{\mbf{P}_{#1}}
\newcommand{\noiseMatrix}[1]{\mbf{W}_{#1}}
\newcommand{\numSubcarrier}{N}
\newcommand{\covMtx}{\mbf{R}}
\newcommand{\precodingVector}{\mbf{p}}
\newcommand{\eigenVectorMax}{\mbf{u}_1}
\IEEEoverridecommandlockouts
\usepackage{cite}
\usepackage{standalone}
\usepackage{tikz}
\usepackage{pgfplots}
\usepgfplotslibrary{polar}
\pgfplotsset{compat=1.17}
\usepackage{amsmath,amssymb,amsfonts}
\usepackage{algorithmic}
\usepackage{graphicx}
\usepackage{textcomp}
\usepackage{epstopdf}
\usepackage{xcolor}
\usepackage{bm}
\graphicspath{{./Figures/}}
\DeclareGraphicsExtensions{.pdf,.jpeg,.png,.eps}
\def\BibTeX{{\rm B\kern-.05em{\sc i\kern-.025em b}\kern-.08em
    T\kern-.1667em\lower.7ex\hbox{E}\kern-.125emX}}
\begin{document}

\title{LPD-Aware Uplink CSI-based 5G NR Downlink Synchronization for Tactical Networks}

\author{\IEEEauthorblockN{Karthik Upadhya, Akshay Jain, Mikko A. Uusitalo}
\IEEEauthorblockA{\textit{Radio Systems Research} \\
\textit{Nokia Bell Labs, Espoo, Finland}\\
\{karthik.upadhya,akshay.2.jain,mikko.uusitalo\}@nokia-bell-labs.com}
\and
\IEEEauthorblockN{Harish Viswanathan}
\IEEEauthorblockA{\textit{Radio Systems Research} \\
\textit{Nokia Bell Labs, Murray Hill, NJ, USA}\\
harish.viswanathan@nokia-bell-labs.com}
}
\maketitle

\begin{abstract}
5G NR is touted to be an attractive candidate for tactical networks owing to its versatility, scalability, and low cost. However, tactical networks need to be stealthy, where an adversary is not able to detect or intercept the tactical communication. In this paper, we investigate the stealthiness of 5G NR by looking at the probability with which an adversary that monitors the downlink synchronization signals can detect the presence of the network. We simulate a single-cell single-eavesdropper scenario and evaluate the probability with which the eavesdropper can detect the synchronization signal block when using either a correlator or an energy detector. We show that this probability is close to $ 100\% $ suggesting that 5G out-of-the-box is not suitable for a tactical network. We then propose utilizing the uplink channel-state-information to beamform the downlink synchronization-signals towards the tactical user-equipment (UE) to lower the eavesdropper detection probability while not compromising the performance of the legitimate tactical UE.
\end{abstract}


\section{Introduction} 
\label{sec:introduction}
Communication has always played a crucial role in battlefields. Effective communication is essential for coordinating troops, providing situational awareness, gathering intelligence, and mounting a rapid response. Access to up-to-date information and enhanced situational awareness can provide an edge in tactical decision-making, often being an important factor in determining the outcome of a tactical scenario. This necessity for a strong communication backbone is becoming more pertinent with modern and future battlefields that are heading towards automation.

Unmanned aerial and ground vehicles require a communication backbone that allows different devices and machines to communicate with each other as well as the command center. Such communications require high data rates and connection densities, trustworthiness, low latencies, and reliability guarantees while also being hard to detect or jam by an adversary. 

To this end, the fifth-generation new radio (5G NR) standard is a promising candidate for providing connectivity in the battlefield since it has been designed to handle a diverse range of use-cases with technologies such as enhanced mobile broadband (eMBB), ultra-reliable low-latency communication (URLLC), and massive machine-type communication (mMTC). For defense applications, this diversity would mean that a single communication standard can allow for high data-rate communications for video surveillance through eMBB, control of unmanned devices through support for URLLC, and support for IoT use-cases, such as perimeter sensing, through mMTC. The added benefit of using a commercial communication standard such as 5G NR is in the reduced platform cost obtained by reusing hardware and IP developed for consumer purposes, as well as exploiting the economies of scale.

On the other hand, it is imperative that 5G NR has a low probability of detection (LPD) and intercept (LPI) and resilience to jamming over all the different phases of communication (initial access, uplink and downlink transmission, control channels etc.) if it has to be deployed as a tactical network by the government or the military. However, downlink (DL) synchronization, which is one of the first steps that a base station (gNB) and user equipment (UE) pair perform to initiate a 5G NR link, has been designed by 3GPP to maximize coverage and reduce complexity and power consumption for a UE that is trying to identify and connect with the gNB\cite{chakrapani2020design,lin2018ss}. Such a design increases the probability of detection (PD) and intercept (PI) and, therefore, is antithetical to the design of a tactical network. 

In this paper, we evaluate the PD in the DL synchronization phase in a 5G NR network with a single-cell, single-user, and single-eavesdropper, and show that the PD is close to $ 100\% $. Since such a high PD is unacceptable for use in a tactical network, we propose a method to reduce the PD through UE-side initial access where the UE sends a proprietary UL transmission to the gNB asking it to initiate the DL synchronization procedure. The gNB obtains the uplink (UL) channel state information (CSI) from the UE's UL transmission and generates the precoding weights for the DL synchronization signal using this CSI. Consequently, this approach reduces the amount of energy that is beamed in the direction of the eavesdropper, which results in a reduction in the PD.

The article is organized as follows: the 5G NR DL synchronization signal structure and the associated challenges in using this approach in a tactical network is presented in Section \ref{sec:stateOfTheArt}. The proposed UL-CSI based DL synchronization method is described in Section \ref{sec:csiBasedDLSynchronization}. The PD performance for both the conventional and proposed methods are presented in Section~\ref{sec:simulationResults}. The paper is then concluded in Section \ref{sec:conclusion}.

\section{5G NR Downlink Synchronization: State of the Art and Limitations}
\label{sec:stateOfTheArt}

The initial access procedure in 5G NR consists of two parts, namely, DL synchronization and UL synchronization. In the first phase, the gNB transmits an always-on DL synchronization signal over the broadcast channel. The UE listens to these DL synchronization signals to extract relevant information from the synchronization signals (SS) and master information block (MIB) and then initiates UL synchronization through the random-access channel (RACH) procedure. Once the UE decodes the MIB, it follows up with the second phase, i.e., UL synchronization.

In addition to being performed when the UE is turned on (i.e., during initial access), these UL and DL synchronization procedures are also performed when the UE in radio resource control (RRC) idle mode becomes active, and during handovers. 

The specifics about the signal transmitted in the DL-synchronization phase have been discussed in Section \ref{subsec:signalStructure}. However, UL synchronization is beyond the scope of this work since the impact on the overall PD and PI from that phase of initial access is smaller than that of DL synchronization. For more details on UL synchronization, we refer the reader to \cite{giordani2016comparative,lin20195g}.

\subsection{Signal Structure} 
\label{subsec:signalStructure}

The DL synchronization phase of the initial access procedure involves broadcasting the synchronization signal block (SSB) which is a combination of the SS and physical broadcast channel (PBCH). The configuration parameters for the SSB at both sub-6 GHz (FR1) and millimeter-wave frequencies (FR2) are provided in Table \ref{table:ssbConfigurationParameters}. 

\begin{table}[htb]
\caption{SSB Configuration Parameters for FR1 and FR2}
\begin{center}
\begin{tabular}{|p{3cm}|p{2cm}|p{2cm}|}
\hline
\textbf{SSB Parameter Names}&\textbf{FR1}&\textbf{FR2} \\ \hline 
Subcarrier Spacing (SCS) & 15 kHz or 30 kHz & 120 kHz or 240 kHz \\ \hline
Number of SS Beams per SS Burst & 4 or 8 & Up to 64  \\ \hline
SSB Type & Case A (15 kHz SCS)  Case B, C1 and C2 (30 kHz SCS) & Case D (120 kHz SCS) or Case E (240 kHz SCS) \\ \hline
\end{tabular}
\label{table:ssbConfigurationParameters}
\end{center}
\end{table}

The main functionality of the SSB is that the UE, by detecting and decoding the SSB, can perform DL synchronization in frequency and time, and acquire critical network information such as the MIB within the PBCH and the physical cell identity (PCI). 

The SS consists of two synchronization signals, i.e., the primary SS (PSS) and secondary SS (SSS) which are binary phase-shift keying (BPSK) modulated m- and Gold-sequences, respectively, with a sequence length of 127 \cite{chakrapani2020design, TS38211, ts38213}. 

%The parameters ${\mathrm{NID1}\in \left\{0,1,\ldots,335\right\}}$ and ${\mathrm{NID2}\in\left\{0,1,2\right\}}$ form the physical cell ID (PCI). The PSS is generated with NID2 and the SSS is generated by using both NID1 and NID2 \cite{TS38211}. Consequently, by detecting the right PSS and SSS sequences, the UE is able to obtain NID1 and NID2 and then evaluate the PCI from Eq. \ref{eqn:pciDefinition}.
%\begin{equation}
%\mathrm{PCI} = 3\times \mathrm{NID1} + \mathrm{NID2}
%\label{eqn:pciDefinition}
%\end{equation}

%One can see that there are 1008 different physical cell identities that can be obtained from 1008 different combinations of the NID1 and NID2. Consequently, there are 1008 possibilities for the SS that can be constructed corresponding to 3 possibilities for the PSS and 336 possibilities for the SSS. 

\begin{figure}[htb] 
\centering
\includegraphics[width=0.75\columnwidth]{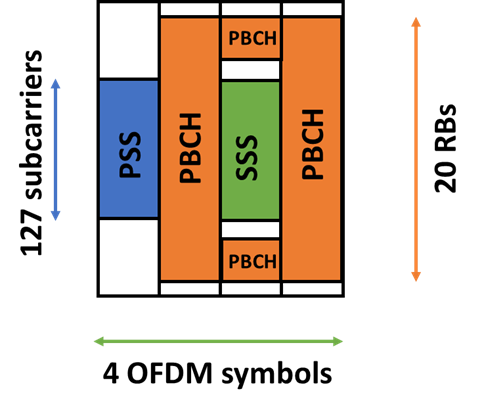}
\caption{SS block time-frequency grid structure }
\label{fig:ssbTimeFrequencyGridStructure}
\end{figure}

Next, the time-frequency grid structure of a SSB is presented in Fig. \ref{fig:ssbTimeFrequencyGridStructure}. It can be seen from this figure that an SSB spans $ 4 $ OFDM symbols in the time domain and $ 20 $ resource blocks (RBs) in the frequency domain. The first symbol contains only the PSS and is used by the UE for detecting the SSB and performing coarse frequency estimation. The third symbol in an SSB contains the SSS surrounded by the PBCH, and the UE utilizes the former to evaluate the reference signal received power (RSRP).
%The UE then utilizes NID1 and NID2 to calculate the PCI, which in turn determines the starting location of the PBCH DMRS in frequency which is given as $\mathrm{mod}(\mathrm{PCI}, 4)$. Subsequently, the UE is able to estimate the channel using the PBCH DMRS and then decode the PBCH which provides it with the information sent in the MIB by the gNB. 

\begin{figure}[htb] 
	\centering
	\includegraphics[width=\columnwidth]{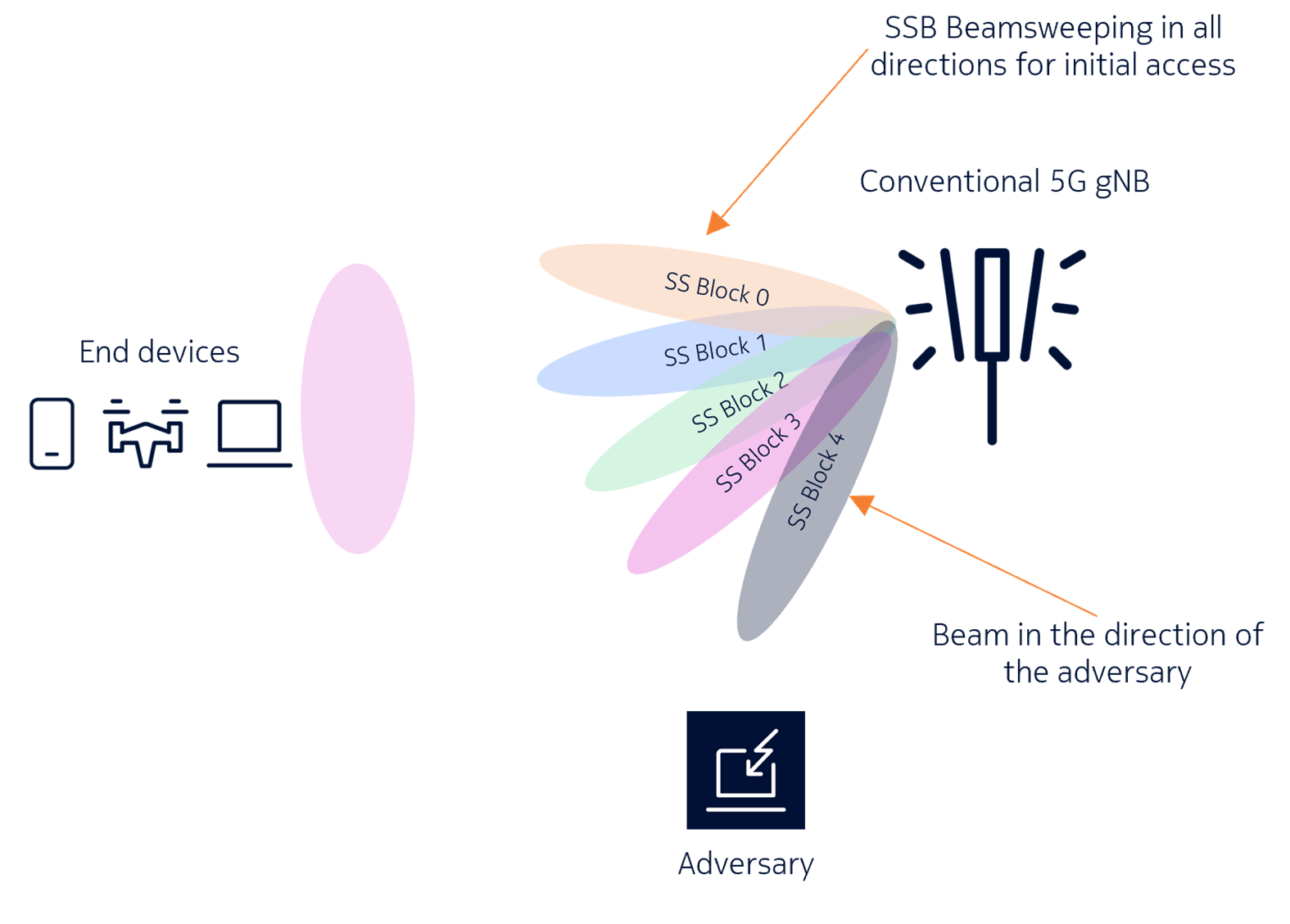}
	\caption{Baseline 5G NR system}
	\label{fig:baseline}
\end{figure}

\begin{figure*}[htb] 
\centering
\includegraphics[trim={0cm 0cm 0cm 0.625cm}, clip, width=0.7\textwidth]{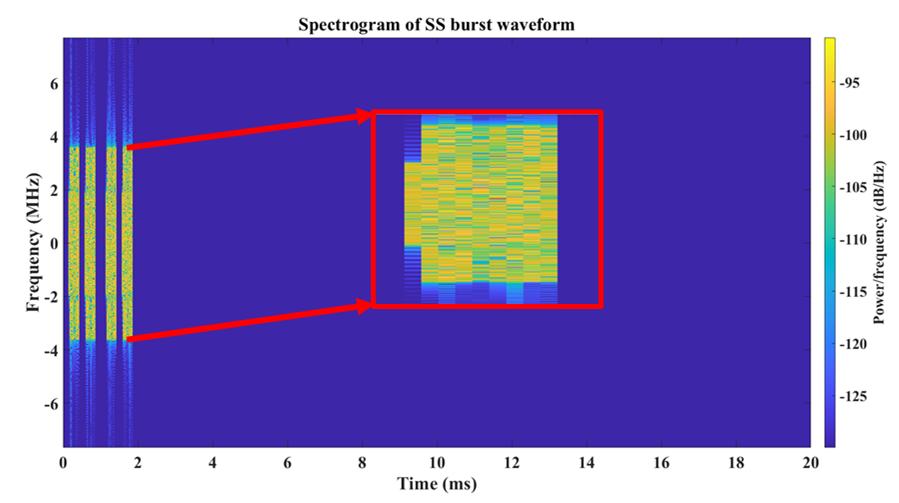}
\caption{Spectrogram of the SSB Burst with 20 ms periodicity for Case B type SSB in FR1 with 8 SSB beams. The figure also includes a zoomed-in image of two SS Blocks. }
\label{fig:SSBSpectrogram}
\end{figure*}

A gNB does not transmit a single SSB, but rather, a burst of multiple SSBs with each SSB transmitted on a different beam such that the entire SSB burst is steered over the entire cell sector \cite{chakrapani2020design, TS38211, ts38213}. This beamsweeping is shown in Fig. \ref{fig:baseline} and is done to utilize the array gain from having multiple antenna ports at the gNB to improve coverage, which is essential especially with massive multiple-input multiple-output (MIMO) or at millimeter-wave frequencies. The number of SSBs in a burst is dependent on the frequency range of operation and is shown in Table \ref{table:ssbConfigurationParameters}, and the structure of a typical SSB burst for the Case B pattern of the SSB in FR1 with 8 SSB beams is shown in Fig. \ref{fig:SSBSpectrogram}. 

%It must also be stated that the SSB might not be located in the center of the band as in LTE \cite{chakrapani2020design, TS38211, ts38213}. Instead, parameters such as the $k_{ssb}$ (location of the  first subcarrier of the SSB with respect to the common resource block \cite{TS38211, ts38213}) and $N_{SSB}^{CRB}$ (this is the common resource block number in which the SSB is positioned and it is usually measured as an `offset to point A' \cite{TS38211, ts38213}) are utilized to determine the location of the SSB, wherein the set of locations is sparse to allow for fast scanning and synchronization.

\subsection{Limitations in Tactical Scenarios} 
\label{sec:DLSynchronizationChallenges}

In commercial 5G NR networks, it is important that the initial access process begins with the gNB transmitting the SSB and the UE performing DL synchronization. This approach enables fast, reliable, and secure initial-access without penalizing UE battery life. Without such an approach, the UE would have to initiate access with its limited transmit power, which limits coverage and consumes battery life in areas where there is no coverage. 

However, for tactical scenarios which need LPD and LPI as stated in Section \ref{sec:introduction}, the DL-first 5G NR initial access process presents some challenges, namely:
\begin{itemize}
	\item DL synchronization utilizes beamsweeping over a grid-of-beams (GoB) to utilize array gain and improve coverage. The GoB typically covers the entire cell-sector. However, as is shown in Fig.~\ref{fig:baseline}, in a tactical scenario that may have eavesdroppers whose objectives are to detect the presence of a tactical network, transmissions in directions other than where the tactical UE is located will increase the PD and PI.
	\item The SSB is an always-on signal and the gNB transmits it even when there are no UEs requiring initial access. These unnecessary transmissions increase the likelihood with which an eavesdropper can detect the presence of the tactical network.
	\item The adversary can easily locate and detect/decode the SSB given that its configuration as well as the possible time and frequency locations of the synchronization signals are standardized and publicly known.   
	\item The PSS and SSS in the SS block are an m-sequence and Gold sequence, respectively, and the method to generate these sequences are specified in 3GPP TS 38.211 \cite{TS38211}. These sequences are also designed with complexity in mind so that a UE can detect these sequences without too much computation. These characteristics pose a significant challenge for tactical 5G NR systems that want to stay hidden.
	\item Transmissions in 5G NR are several PRBs wide and their location in frequency is also defined with PRB-level granularity. Consequently, frequency hopping in 5G NR is coarse in comparison to what is typically done in tactical systems where frequency hopping is done with transmissions that are a few kHz wide and hopped with a fine granularity \cite{elmasry2021hiding}.
\end{itemize}

\begin{figure}[htb] 
	\centering
	\resizebox{\columnwidth}{!}{\input{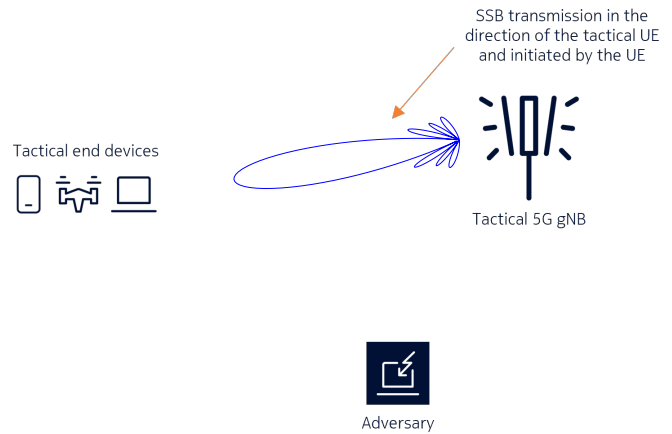}}
	\caption{5G NR system with CSI-based DL beamforming, where the gNB adapts the DL beam based on the UL CSI and sends the SSB beam in the direction of the UE}
	\label{fig:enhanced}
\end{figure}

\section{Uplink CSI-based DL Synchronization}
\label{sec:csiBasedDLSynchronization}

In this section, we propose replacing the always-on SSB transmission and GoB-based beamsweeping with an on-demand SSB transmission that is adaptively beamformed based on measured UL CSI. Both these schemes have the potential to substantially reduce the PD during initial access. The on-demand SSB transmission reduces unwanted transmissions whereas the UL-CSI-based beamforming forces transmission only in the direction of the UE, as illustrated in Fig.~\ref{fig:enhanced}.

On-demand SSB transmission is performed through UE-sided initial access by the UE transmitting a proprietary signal in the UL. The proprietary signal includes a random pilot, known only to the tactical UE and the gNB, and a payload that contains parameters such as the UL transmit power. The gNB then detects this transmission and estimates the UL CSI from the received signal. The estimated CSI is then used to compute the precoding matrix for the DL synchronization signal. Finally, the gNB transmits the precoded DL synchronization signal to the tactical UE. 

The proposed UE-sided initial access approach can be implemented in several ways. One such implementation, that does not require any modifications to the hardware of a COTS UE, is presented in \cite{ourpaper} and involves attaching a low-cost software-defined radio transmitter and receiver to the UE and the gNB, respectively.

A mathematical description of the CSI estimation and precoding is described as follows\footnote{Note that the details of the UL transmitted waveform and the detection algorithm at the gNB is out of the scope of this paper. This is because the PD of an eavesdropper is dominated by the DL synchronization phase since the SSB is beamformed and transmitted at a much higher transmit power than the UL UE-sided initial access transmission.}. We consider a single-cell single-UE system\footnote{The approach can easily be generalized for a multi-cell system with multiple UEs. We restrict ourselves to the single-cell single-UE setup for the ease of exposition.} with the gNB having $ \numBsAntenna $ antennas and the UE having $ \numUeAntenna $ antennas. The UL channel between the UE and the gNB on the $ \subcarrierIdx\rth $ subcarrier is denoted as $ \channelMtx{\subcarrierIdx} \in \complex^{\numBsAntenna\times\numUeAntenna} $.  

For the purposes of estimating the CSI, we only consider the received signal corresponding to the UL transmission. The received signal in the UL in the $ \subcarrierIdx\rth $ subcarrier $ {\receivedSignal{\subcarrierIdx}\in\complex^{\numBsAntenna\times\numUeAntenna}} $ can then be written as 
\begin{align}
	\receivedSignal{\subcarrierIdx} 
	=
	\sqrt{\ulTxPower} 
	\channelMtx{\subcarrierIdx}	
	\ulPilotSignal{\subcarrierIdx}
	+
	\noiseMatrix{\subcarrierIdx} \;,
	\label{eqn:receivedSignal}
\end{align}
where $ \ulTxPower $ is the UL transmit power and $ \ulPilotSignal{\subcarrierIdx} \in \complex^{\numUeAntenna\times\numUeAntenna} $ is the pilot matrix transmitted by the UE. The pilot matrix is assumed to be unitary, i.e., $ \ulPilotSignal{\subcarrierIdx}^H\ulPilotSignal{\subcarrierIdx} = \ulPilotSignal{\subcarrierIdx}\ulPilotSignal{\subcarrierIdx}^H = \eye_{\numUeAntenna} $ where $ (\cdot)^H $ represents the Hermitian transpose of a matrix.

Since no a priori information of the channel statistics can be assumed at the gNB, we utilize the least-squares estimate of the channel. This estimate can be obtained from \eqref{eqn:receivedSignal} as
\begin{align}
	\channelEstimate{\subcarrierIdx} 
	= 
	\frac{1}{\sqrt{\ulTxPower}}
	\receivedSignal{\subcarrierIdx}
	\ulPilotSignal{\subcarrierIdx}^H
	=
	\channelMtx{\subcarrierIdx}
	+
	\frac{1}{\sqrt{\ulTxPower}}
	\noiseMatrix{\subcarrierIdx}
	\ulPilotSignal{\subcarrierIdx}^H \;.
	\label{eqn:channelEstimate}
\end{align}
Next, the precoding vector for transmitting the SSB is obtained from the channel estimate through eigenbeamforming. Specifically, the gNB computes the covariance matrix $ \covMtx \in \complex^{\numBsAntenna\times\numBsAntenna}$ from the channel estimate $ \channelEstimate{\subcarrierIdx} $ as
\begin{align}
	\covMtx
	=
	\frac{1}{\numSubcarrier}
	\sum\limits_{\subcarrierIdx=0}^{\numSubcarrier-1}
	\channelEstimate{\subcarrierIdx}
	\channelEstimate{\subcarrierIdx}^H \;.
\end{align}
Then, the precoding vector $ \precodingVector \in \complex^\numBsAntenna $ is given as $ \precodingVector = \eigenVectorMax^* $, where $ \eigenVectorMax $ is the eigenvector of the covariance matrix $ \covMtx $ corresponding to the largest eigenvalue and $ (\cdot)^* $ represents the conjugate of a vector. 

\section{Simulation Scenario, Analysis and Discussions} 
\label{sec:simulationResults}
In this section, we simulate the baseline 5G NR system (c.f. Fig.~\ref{fig:baseline}) and the proposed UL-CSI-based DL synchronization transmission (c.f. Fig.~\ref{fig:enhanced}) and evaluate the PD in both cases. Table \ref{tab:simparams} presents the simulation scenario and its related parameters. Note that for the CSI-based beamforming, we use the exact CSI over a single slot ($ 0.5 $ ms at $ 30 $ kHz subcarrier spacing) for evaluating the covariance matrix $ \covMtx $ and the precoding matrix. 

In addition, for the proposed approach, only the sector with the highest RSRP to the UE transmits the DL synchronization signal. This is because the gNBs can evaluate the RSRP from the UL received signal and determine the sector where the UE is located. This is in contrast to the baseline approach where all the cell sectors transmit the SSB in the DL.

\begin{table}[htb]
\caption{Simulation Parameters}
\begin{center}
\begin{tabular}{|p{4.5cm}|p{3.5cm}|}
\hline
\textbf{Parameter Name}&\textbf{Value} \\ \hline 
Number of cell sectors & $ 3 $ \\ \hline
Sector width & $ 120^\circ $ \\ \hline
Number of UEs & $ 1 $ \\ \hline
UE antenna array configuration $ (\mathrm{rows} \times \mathrm{columns} \times \mathrm{polarization}) $ & $ 2 \times 1 \times 2 $ \\ \hline
UE location & Randomly distributed over the three cell sectors\\ \hline
gNB antenna array configuration $ (\mathrm{rows} \times \mathrm{columns} \times \mathrm{polarization}) $  & $ 4 \times 2 \times 2 $ and $ 8 \times 8 \times 2 $ \\ \hline
Subcarrier spacing & $ 30 $ kHz \\ \hline
SSB transmission period & $ 20 $ ms \\ \hline
Number of eavesdroppers & $ 1 $ \\ \hline
Eavesdropper antenna array configuration $ (\mathrm{rows} \times \mathrm{columns} \times \mathrm{polarization}) $ & $ 2 \times 1 \times 2 $\\ \hline
Eavesdropper location & Randomly distributed over the three cell sectors \\ \hline
Eavesdropper observation bandwidth & $ 15.36 $ MHz \\ \hline
Eavesdropper observation time & $ 25 $ ms \\ \hline
gNB transmit power & $ 28 $ dBm for the $ 4 \times 2 \times 2 $ array and $ 19 $ dBm for the $ 8 \times 8 \times 2 $ array \\ \hline
Channel model & 3GPP $ 38.901 $ UMi \\ \hline
Carrier frequency & $ 3.5 $ GHz \\ \hline
Cell inter-site distance & $ 200 $ m \\ \hline
\end{tabular}
\label{tab:simparams}
\end{center}
\end{table}

\begin{figure*}	
	\begin{minipage}[t]{0.48\textwidth}
		\centering
		\includegraphics[width=\columnwidth]{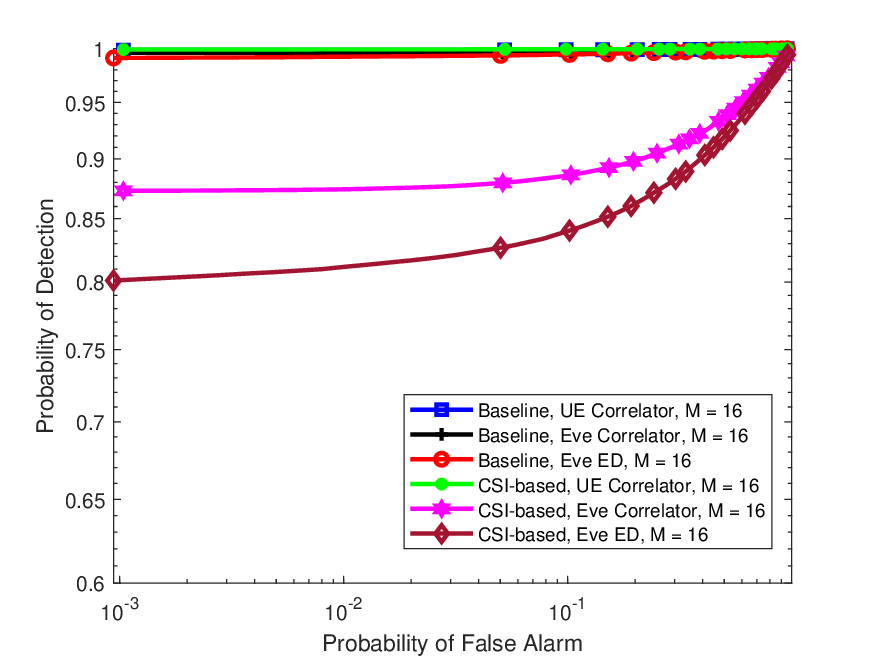}
		\caption{ROC for the baseline and UL-CSI-based DL synchronization with $ \numBsAntenna = 16 $ antennas.}
		\label{fig:ROC_M=16}
	\end{minipage}
	\hspace{\stretch{3}}
	\begin{minipage}[t]{0.48\textwidth}
		\centering
		\includegraphics[width=\columnwidth]{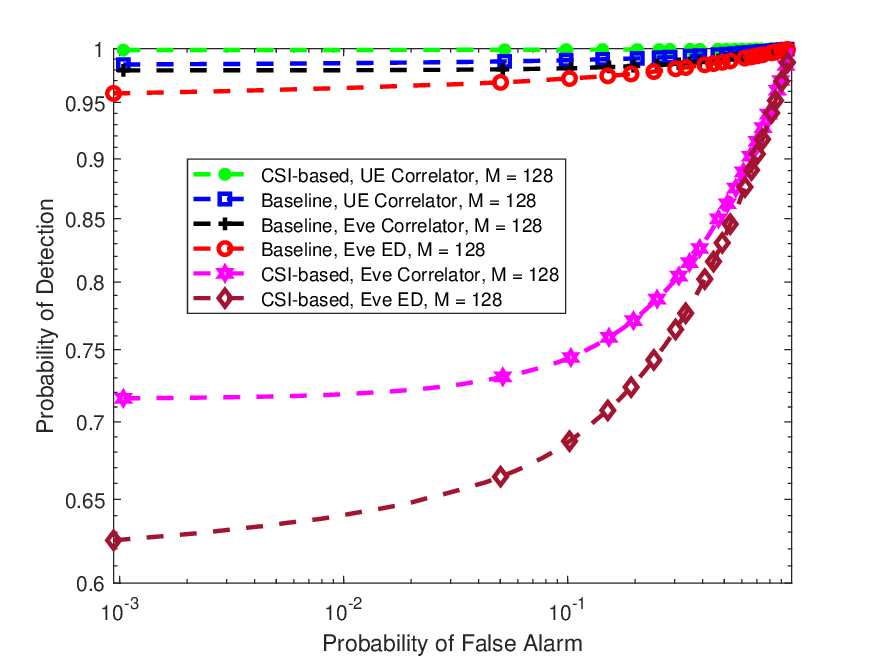}
		\caption{ROC for the baseline and UL-CSI-based DL synchronization with $ \numBsAntenna = 128 $ antennas.}
		\label{fig:ROC_M=128}
	\end{minipage}	
	\vspace{-0.25cm}
\end{figure*}

The gNB transmit power is set such that the cell-edge RSRP/SNR at the UE is similar to that of real-world deployments. This benchmarking is done using a system level simulator with the 3GPP 38.901-UMi channel model which results in the gNB transmit power being set to $ 28 $ dBm and $ 19 $~dBm for the array with $ 16 $ and $ 128 $ antenna ports, respectively, for the bandwidth of $ 20 $ PRBs (corresponding to the width of an SSB) and an inter-site distance of $ 200 $ m. Note that the transmit power is only for transmitting the SSB and not the full carrier.

We then evaluate the PD at the eavesdropper with both an energy detector and a correlator which are introduced as follows:
\begin{itemize}
	\item \textbf{Energy Detector:} This detector assumes that the eavesdropper does not have the knowledge of the DL synchronization signal structure nor its location in time or frequency. Instead, it scans across the observation bandwidth and time window specified in Table \ref{tab:simparams}. The scanning process involves the UE utilizing a sliding window of width $ 20 $ PRBs in frequency and $ 4 $ OFDM symbols in time (assuming a subcarrier spacing same as that of the tactical system) over the observation time and bandwidth. The detector computes the received energy in each position of the sliding window and declares the existence of the DL synchronization signal when the energy at any one of these positions exceeds a threshold. The threshold is varied so as to plot the receiver operating characteristic (ROC).
	\item \textbf{Correlator:} Given that the PSS and SSS are standardized sequences, an eavesdropper with knowledge of these sequences can implement a correlator to detect the presence of the DL synchronization signal. 
	
	However, similar to the energy detector, this receiver does not assume to know the location of the PSS or SSS in time or frequency and instead, scans across the observation bandwidth and time window. For each value of time and carrier frequency offset (the former is at sample spacing and the latter is at RB spacing), the receiver correlates the received signal with the PSS and SSS and declares the existence of the DL synchronization signal when the output of the correlator exceeds a threshold at any combination of time or carrier frequency offset in the window. Here again, the threshold is varied so as to plot the ROC.
	
	This detector is expected to perform better than the energy detector since it can utilize the coherent processing gain from correlation.
\end{itemize}

Next, in Figs. \ref{fig:ROC_M=16} and \ref{fig:ROC_M=128}, the ROC is plotted for the UE and the eavesdropper with $ \numBsAntenna=16 $ and $ \numBsAntenna=128 $ antennas, respectively. The UE uses a correlator for the detection and the eavesdropper uses an energy detector and a correlator. 

In both cases, it is clear that the baseline scenario results in a PD of almost $ 100 $\% for the eavesdropper. This result substantiates our earlier claim that the 5G NR DL synchronization is optimized for coverage. Specifically, because to the grid-of-beams approach, the beams are not well designed for spatial selectivity but rather are designed to offer coverage. This allows the energy detector to detect the DL synchronization signal transmissions. Moreover, the PSS and SSS sequences are of length 127 and offer an approximately $ 20 $ dB coherent processing gain when used with a correlator and consequently, the correlator performs better than the energy detector. This is true in both Figs. \ref{fig:ROC_M=16} and \ref{fig:ROC_M=128}, albeit not as pronounced in the former plot, since the high gNB transmit power and coverage characteristics of the SSB transmission ensures good detection capabilities for the energy detector as well.

When UL-CSI-based beamforming is employed, there is a drop in the PD for the eavesdropper without any impact on the PD for the UE. The gains can be attributed to two different reasons
\begin{itemize}
	\item In the baseline scenario, all three sectors are transmitting the DL synchronization signal, since the gNBs cannot estimate the location of the UE or the propagation characteristics to it. However, for the proposed approach, only the gNB corresponding to the sector with the highest RSRP transmits the DL synchronization signal, whereas the gNBs in the remaining two cell sectors are silent.
	\item The eigenbeamforming approach increases spatial selectivity of the DL transmission.
\end{itemize} 
The reduction in PD for the eavesdropper is higher with $ {M=128} $ owing to the higher spatial selectivity that results in a lower likelihood of energy being radiated in the direction of the eavesdropper. Increasing the number of antennas further will further reduce the PD for the eavesdropper.

\section{Conclusions} 
\label{sec:conclusion}

In this paper, we assess the suitability of 3GPP based 5G NR for tactical networks focusing on the LPI/LPD characteristics. 5G NR initial access begins with downlink synchronization signals designed to facilitate UEs to discover network gNBs and synchronize to them. Specifically, the DL synchronization phase involves the gNB transmitting the SSB in all directions at a high transmit power to ensure cell-edge coverage. Based on simulations, it was observed that the conventional approach results in a PD that is close to $100\%$ if the gNB transmits the SSB at a power that allows it to provide coverage at the cell edge.

The high PD is attributed to: (i) fat beams for DL synchronization that overlap to ensure coverage and do not offer spatial isolation even when the right transmit beam is selected (ii) high transmit power that is independent of the UE location and (iii) known/fixed sequences in the SSB that allow an eavesdropper to obtain coherent processing gain with a correlator. 

Subsequently, we simulated the proposed UL-CSI-based approach and found that the PD at the eavesdropper is significantly lower than the baseline without any penalty to the UE's detection performance. The PD was seen to reduce when increasing the number of gNB antennas, indicating that massive MIMO is useful to obtain LPD in tactical networks even in initial access. 

Further reduction in the PD can be obtained by combining the proposed UL-CSI-based approach with DL power control in \cite{ourpaper}. This is left as future work.

\bibliographystyle{IEEEbib}
\bibliography{IEEEabrv,references}

\end{document}